\documentclass[a4paper,twoside]{article}
\newcommand{\affil}[1]{$^{\rm #1}$}
\usepackage[authoryear]{natbib}
\bibpunct{(}{)}{;}{a}{}{,}
\usepackage{graphicx}
\date{Publications of the Astronomical Society of Australia\\
Paper accepted on the 4th of July, 2008} 
\title{\large\bf\flushleft 
Populating the galaxy velocity dispersion -- supermassive black hole mass
diagram: A catalogue of ($M_{\rm bh}, \sigma$) values}
\author{\parbox{\textwidth}{\flushleft
\vspace{-0.5cm}
{\it
Alister W.\ Graham\affil{A,}\affil{B} 
}\\
\vspace{0.4cm}
{\small 
  \affil{A}\, Centre for Astrophysics and Supercomputing, Swinburne
  University of Technology, Hawthorn, Victoria 3122, Australia.\\
  \affil{B}\,Email: AGraham@astro.swin.edu.au \\
}}}
\begin{document}
\twocolumn[
\begin{changemargin}{.8cm}{.5cm}
\begin{minipage}{.9\textwidth}
\vspace{-1cm}
\maketitle

\small{\bf
  Abstract: An updated catalogue of 76 galaxies with direct
  supermassive black hole mass measurements ($M_{\rm bh}$) plus, when
  available, their host bulge's central velocity dispersion ($\sigma_0$) is
  provided.  Fifty of these mass measurements are considered reliable, while
  the others remain somewhat uncertain at this time.  An additional eight
  stellar systems, including one stellar cluster and three globular clusters,
  are listed as hosting potential intermediate mass black holes $< 10^6
  M_{\odot}$.

  With this larger data set, the demographics within the $M_{\rm
    bh}$-$\sigma_0$ diagram are briefly explored.  Many barred galaxies are
  shown to be offset from the $M_{\rm bh}$-$\sigma_0$ relation defined by the
  non-barred galaxies, in the sense that their velocity dispersions are too
  high.  Furthermore, including 88 AGN with black hole mass estimates from
  reverberation mapping studies, we speculate that barred AGN may follow this
  same general trend.  We also show that some AGN with $\sigma_0 < 100$ km
  s$^{-1}$ tend to reside up to 0.6 dex above the {\it barless} $M_{\rm
    bh}$-$\sigma_0$ relation.  Finally, it is shown that ``core galaxies''
  appear not to define an additional subdivision of the $M_{\rm
    bh}$-$\sigma_0$ diagram, although improved methods for measuring
  $\sigma_0$-values may be valuable.
}


\medskip
\medskip
\end{minipage}
\end{changemargin}
]
\small

\section{Introduction}


Scaling relations between the intrinsic properties of galaxies provide clues to
the physical mechanisms which operate within these systems.  In general, the
tighter a relation is, i.e.\ the less scatter it has, the more fundamental the
relation is expected to be.  Therefore, it is perhaps not surprising that
there has been a huge interest in the $M_{\rm bh}$-$\sigma$ relation
(Ferrarese \& Merritt 2000; Gebhardt et al.\ 2000) which was reported to have
very little or no intrinsic scatter.  In addition to providing an indirect
means to measure supermassive black hole (SMBH) masses in many galaxies, the
$M_{\rm bh}$-$\sigma$ relation (Merritt \& Ferrarese 2001a; Tremaine et al.\
2002; Ferrarese \& Ford 2005; Novak et al.\ 2006), along with the equally
strong $M_{\rm bh}$-$n$ relation (Graham \& Driver 2007a) and $M_{\rm bh}$-$L$
relation (Kormendy \& Richstone 1995; Magorrian et al.\ 1998; 
McLure \& Dunlop 2002; Marconi \& Hunt 2003; Graham 2007), provides 
insight into the joint formation process of SMBHs and their host bulges.  Such
relations can also be applied to volume-limited galaxy samples, providing an
estimate of the SMBH mass density in the Universe (e.g., Salucci et al.\ 1999;
Graham \& Driver 2007b, and references therein).

As the number of galaxies with direct SMBH mass measurements has increased, it
has become possible to explore the demographics of the SMBH population within
the $M_{\rm bh}$-$\sigma$ diagram.  Rather than delineating a single line,
Graham (2008) and Hu (2008) have revealed a tendency for SMBHs in barred
galaxies, or perhaps equivalently pseudobulges, to reside below the $M_{\rm
  bh}$-$\sigma$ relation defined by non-barred galaxies.  In addition, Hu
(2008) has noted that there may be a third subdivision in the $M_{\rm
  bh}$-$\sigma$ diagram such that ``core galaxies'' 
(Ferrarese et al.\ 1994; Faber et al.\ 1997; Trujillo et al.\ 2004) 
define a steeper relation
than non-core galaxies.  Such departures from a single unifying expression
offer the promise of further valuable clues into the coevolution of galaxies
and the million to billion solar mass black holes which reside at their
centres.

This paper presents the largest sample of galaxies for which direct SMBH mass
estimates are available.  While the structure within the updated $M_{\rm
  bh}$-$\sigma$ diagram is explored here, it is additionally hoped that this
database will be a helpful resource, or rather stepping stone, for future 
investigations.

%

\section{$M_{\rm bh}$ versus $\sigma_0$}

\begin{figure*}[ht]
\begin{center}
\includegraphics[angle=270,scale=0.66]{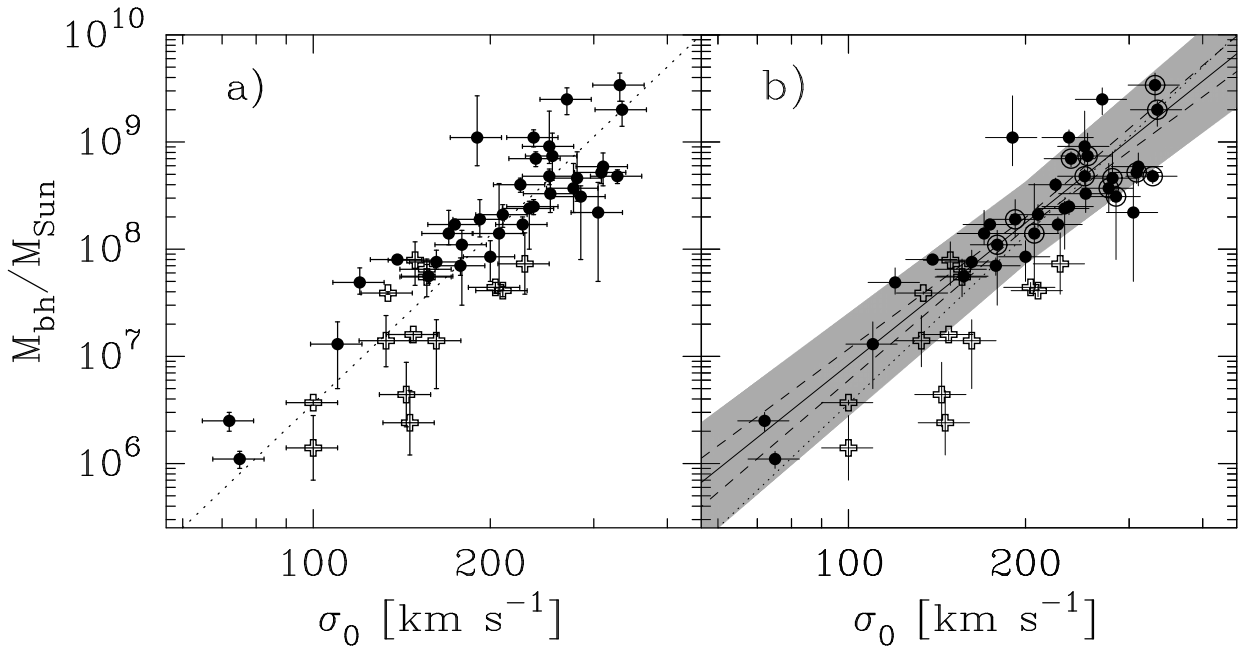}
\caption{
50 galaxies in the $M_{\rm bh}$-$\sigma_0$ diagram (see Table~\ref{Tab1}). 
The 14 barred galaxies are denoted by the crosses. 
Known ``core galaxies'' have been circled in panel b).
The solid line is the optimal linear regression to the non-barred 
galaxies, as given by Eq.~\ref{Eq_BCES}, while the dashed lines delineate 
the 1$\sigma$ uncertainty for this relation.
The shaded area extends this boundary by 0.33 dex in the $\log M_{\rm bh}$ 
direction. 
The dotted line is the linear regression to all 50 data points.
}
\label{Fig1}
\end{center}
\end{figure*}

\subsection{The Data}

Ferrarese \& Ford (2005) presented a highly useful list of 38 galaxies 
for which SMBH mass estimates had been obtained from resolved dynamical
studies. 
Scouring the literature, one finds that this number has doubled over
the past three years.  While some galaxies have most likely been
inadvertently overlooked, Tables~\ref{Tab1} and \ref{Tab2} are believed to
represent the most complete sample of galaxies with direct SMBH mass estimates
published to date.  The reference for each SMBH mass is provided in the final
column of each table.  A total of 50 galaxies are listed in Table~\ref{Tab1}.  They
are considered to have reasonably reliable measurements of their SMBH mass.
The second table contains almost three dozen stellar systems whose SMBH masses are not
yet secure, for the reasons noted in Table~\ref{Tab2}.  It is of course hoped
that in the near future many of these galaxies will migrate into
Table~\ref{Tab1}.

When this paper's adopted distance to a given galaxy differed from the
distance used in the paper which derived the SMBH mass, the mass has been
rescaled here to the new distance.  The adopted distances are listed in
Tables~\ref{Tab1} and \ref{Tab2} along with a reference to the new distance.

The basic morphological Hubble type has been taken from
NED\footnote{http://nedwww.ipac.caltech.edu/}, with the exception that the
galaxies noted in Graham (2008) to be barred are labelled as such, as is
NGC~2639 (M\'arquez et al.\ 1999).  In addition, following Graham \& Driver
(2007a), early-type galaxies with discs are labelled as lenticular (S0) rather
than elliptical (E).

Many giant elliptical galaxies are known to possess partially depleted stellar
cores relative to the inward extrapolation of their outer light profile (e.g.,
Kormendy 1985; Lauer 1985).  A long-standing idea for the production of such
cores is from the gravitational scouring and ejection of stars by SMBHs
(Begelman, Blandford, \& Rees 1980; Ebisuzaki, Makino, \& Okumura 1991; Makino
\& Ebisuzaki 1996; Quinlan 1996; Quinlan \& Hernquist 1997).  Typical central deficits
in stellar mass are on the order of the mass of the central SMBH (Graham 2004;
Ferrarese et al.\ 2006, their Section~5.2; Merritt 2006).  The gravitational
recoil of the final, merged SMBH may also contribute to this reduction of the
central stellar density (e.g., Gualandris \& Merritt 2008).  One may therefore
expect the so-called ``core galaxies'', formed via dry merger events, to display a
different distribution in the $M_{\rm bh}$-$\sigma_0$ diagram.
Whether or not a galaxy contains a partially-depleted core is noted in
Tables~\ref{Tab1} and \ref{Tab2} using the identifications in Faber et al.\ (1997), 
Quillen et al.\ (2000), Ravindranath et al.\ (2001) and Rest et al.\ (2001).

As an inspection of HyperLeda\footnote{http://leda.univ-lyon1.fr/} will
reveal, the published central velocity dispersion, $\sigma_0$ of many galaxies
can vary quite substantially.  Most galaxies do not have flat velocity
dispersion profiles, and so the radius within which one measures the
velocity dispersion is an issue\footnote{In addition, for small apertures, the seeing
conditions can influence the measurements even when the sampling radius
remains unchanged.}.  
Jorgensen, Franx \& Kjaergaard (1995) provide a correction from $\sigma _0$ to
$\sigma_{\rm e}$, the luminosity-weighted velocity dispersion within one
effective radius $R_{\rm e}$.  It does however assume that the same normalised
velocity dispersion profile exists for all galaxies.  Potential, and indeed
expected, systematic changes in the velocity dispersion profile shape with
host bulge magnitude are therefore ignored by this adjustment.  Rather than try and determine
which value is the most appropriate, this paper has effectively placed its
trust in the averaging process employed by HyperLeda and simply uses the
(February 2008) HyperLeda-supplied central velocity dispersions, $\sigma_0$.

\subsection{The Diagram}

Figure~\ref{Fig1} presents the SMBH masses versus the central velocity
dispersions for the 50 galaxies listed in Table~\ref{Tab1}.

\subsubsection{(Non-)Barred galaxies}

Galaxies known to possess a bar have been designated with a cross in
Figure~\ref{Fig1}.  As observed in Graham (2008, his Figure~5), many barred
galaxies display a tendency to reside below the $M_{\rm bh}$-$\sigma_0$
relation defined by the non-barred galaxies.  A similar behavior was
identified by Hu (2008) for SMBHs deemed to reside in ``pseudobulges''.
It is important to realise that the claim is not that {\it all} barred
galaxies are offset in this diagram, only that some are --- perhaps due to the
streaming motions of their stars influencing the measured velocity dispersion
of the host bulge.

Using the (symmetrical) bisector linear regression routine BCES from Akritas
\& Bershady (1996), and assigning a 10 per cent uncertainty to the Hyperleda
velocity dispersions, for the 36 non-barred galaxies one obtains the relation
\begin{equation}
\log(M_{\rm bh}/M_{\odot})  = (8.25\pm0.05) + (4.39\pm0.32)\log 
[\sigma_0/200\, {\rm km\, s}^{-1}].
\label{Eq_BCES}
\end{equation}
The slope is 4.28 and 4.58 when using an uncertainty of 5 and 15 per cent 
for the velocity dispersion, respectively. 
Although this expression was not obtained by minimising the 
scatter in the $\log M_{\rm bh}$ direction, the {\it total} 
r.m.s.\ scatter in this direction is 0.33 dex.

Using all 50 galaxies, and a 10 per cent uncertainty on the velocity
dispersion, a bisector linear regression gives 
$\log(M_{\rm bh}/M_{\odot})  = (8.13\pm0.06) + (5.22\pm0.40)\log 
[\sigma_0/200\, {\rm km\, s}^{-1}]$.

\subsubsection{Core galaxies}

Hu (2008) reveals that ``core galaxies'' may have a steeper slope in the
$M_{\rm bh}$-$\sigma_0$ diagram than galaxies without partially depleted cores.
This is interesting because it may reflect the different formation history of
the galaxies involved.  Hu notes, however, that the different behavior only
appears when using the velocity dispersions corrected to $R_{e/8}$ via the
prescription given by Jorgensen et al.\ (1995).  The difference is not
evident when using the velocity dispersions within $R_e$ from Tremaine et al.\
(2002).  This mixed result was also evident in the Figures of Wyithe (2006a).
In Figure~\ref{Fig1}b, using the central velocity dispersions from HyperLeda,
and without applying the formula from Jorgensen et al.\ (1995), no obvious
difference to the relation defined by the core and non-core galaxies is
apparent.

Given that the Luminosity-$\sigma_0^{\alpha}$ relation has an exponent $\alpha
\sim 4$ for luminous elliptical galaxies (Faber \& Jackson 1976), but $\alpha
\sim 2$ for dwarf elliptical galaxies (e.g.\ de Rijcke et al.\ 2005;
Matkovi\'c \& Guzm\'an 2005, and references therein), then, as noted in Graham
\& Driver (2007a, their section~3.2), if the $M_{\rm bh}$-Luminosity relation
is linear (Graham 2007) one would expect the $M_{\rm bh}$-$\sigma_0$ relation
to have two different slopes.  While ``core'' galaxies occupy the massive-end
of this diagram, neither they nor the other big elliptical galaxies appear to
define a different (steeper) relation to the ``non-core galaxies''.  The
answer may be due to the prevalence of disc galaxy bulges, rather than dwarf
elliptical galaxies, at the low mass end of the $M_{\rm bh}$-$\sigma_0$
diagram, and it is concluded that an increased galaxy sample with reliable
black hole mass measurements and velocity dispersions would be beneficial in
resolving this issue.

\subsubsection{Active galaxies}

Feedback from Active Galactic Nuclei (AGN) has long been proposed as a
mechanism to curtail both SMBH growth and quench star formation in the host
bulge (Begelman, de Kool \& Sikora 1991; Silk \& Rees 1998; Fabian 1999;
Benson et al.\ 2003; Begelman \& Nath 2005).  This popular idea has been
implemented in semi-analytical simulations of galaxies to shut off stellar
growth in massive elliptical galaxies and explain both the $M_{\rm
  bh}$-$\sigma_0$ relation and the exponential decline at the bright end of the
galaxy luminosity function (Granato et al.\ 2004; Bower et al.\ 2006; Croton
et al.\ 2006).



In spite of AGN clearly signalling the presence of SMBHs, with the exception of
NGC~4395 and Pox~52, only galaxies with direct dynamical measurements of material orbiting
around their central black hole have been tabulated here.  That is, 
galaxies with active nuclei --- whose black holes are thus currently under
construction at some level --- have not been included.  Reverberation mapping
estimates of SMBH masses do however exist for an increasing number of such galaxies, 
although the relatively larger uncertainty on their SMBH masses is not so desirable. 

\begin{figure}[ht]
\begin{center}
\includegraphics[angle=270,scale=0.54]{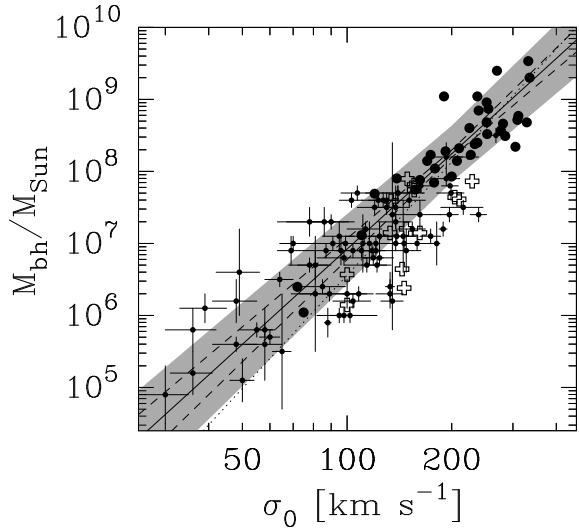}
\caption{
Similar to Figure~\ref{Fig1}b except that the 88 AGN (small points) 
from Greene \& Ho (2006) have been added.
}
\label{Fig2}
\end{center}
\end{figure}

From a sample of 15 Seyfert 1 galaxies, Barth, Greene \& Ho (2005) reported 
that they followed the same $M_{\rm bh}$-$\sigma$ relation as defined by the
local inactive sample from Tremaine et al.\ (2002). 
In contradiction to this, Wyithe (2006a,b) subsequently argued that a fraction
resided above the standard $M_{\rm bh}$-$\sigma_0$ relation, 
evident over the velocity dispersion interval from $\sim$30 to $\sim$90 km
s$^{-1}$ (see also Zhang et al.\ 2008 who used Type II AGN). 
With an increased sample of 88 Type I AGN, and updated SMBH mass estimates, Greene \& Ho (2006) 
have noted that there is indeed some evidence of a flatter slope to the
$M_{\rm bh}$-$\sigma$ relation at the low black hole mass end of the
distribution.  This can be seen in Figure~\ref{Fig2} where these 88 AGN have
been included.  If not due to selection biases or over-estimated SMBH masses,
this result may then signal an additional (third, after the barred galaxies)
zone in the $M_{\rm bh}$-$\sigma$ plane.  

Also evident, but previously unrecognised, is the overlap of some AGN with the
barred galaxies that deviate from the barless $M_{\rm bh}$-$\sigma$ relation.
It would be of interest to identify if the AGN which fall below the barless
$M_{\rm bh}$-$\sigma$ relation also have bars, and it is speculated here that
they probably do.  This is under investigation in Graham \& Li (2008, in
prep.).  
%
%

%
%

\section{Outlook}

\begin{figure}[ht]
\begin{center}
\includegraphics[angle=270,scale=0.54]{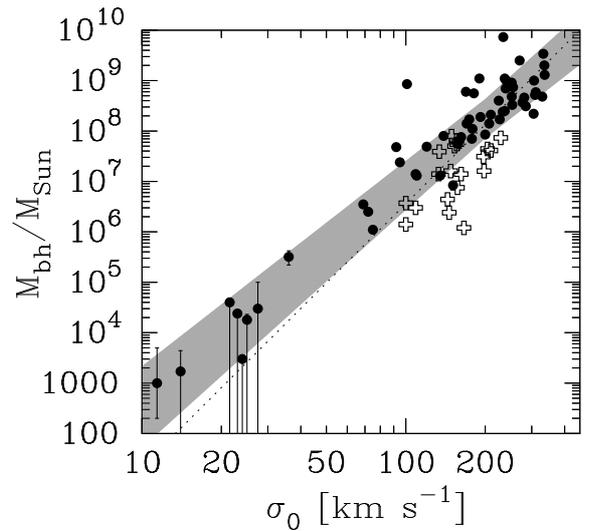}
\caption{
Sixty nine galaxies with both SMBH mass estimates and $\sigma_0$ values, 
plus 8 stellar systems with IMBH mass
estimates (taken from Tables~\ref{Tab1} and \ref{Tab2}). 
The 21 barred galaxies are denoted by the crosses. 
For reference, the shaded area and dotted line is the same as that shown in Figure~\ref{Fig1}.
}
\label{Fig4}
\end{center}
\end{figure}

With the increasing spatial resolution available from current and upcoming
instruments, the number of SMBHs with resolved spheres-of-influence (Merritt
\& Ferrarese 2001b) is set to increase.  Indeed, the community anxiously await
the measurements of SMBH masses in some twenty galaxies from the combination
of SAURON/WHT and OASIS/WHT data (Capellari et al.\ 2008, priv.\ comm.).  The
$M_{\rm bh}$-$\sigma_0$ diagram shown in Figure~\ref{Fig4} is similar to
Figure~\ref{Fig1} except that the SMBH data from both Tables~\ref{Tab1} and
\ref{Tab2} are shown.  While one can see that the inclusion of the additional
(less secure) data has increased the scatter, due no doubt to the greater
uncertainties on these SMBH masses, many barred galaxies still display a
tendency to reside beneath the barless relation established previously
(Equation~\ref{Eq_BCES}).  While the intermediate mass black holes (IMBHs)
appear to follow the barless $M_{\rm bh}$-$\sigma_0$ relation defined by the
more massive systems, it is noted that most of the IMBH masses are not yet
securely established and they may in fact not exist at all --- as noted in
their parent papers.

At present, for most galaxies only an upper-limit on their SMBH mass exists
(e.g., Beifiori et al.\ 2008).  (Active Optics)-enhanced integral field
spectrograph data from instrument/telescope combinations such as NIFS/Gemini,
OSIRIS/Keck, SINFONI/VLT, LUCIFER/LBT and ATLANTIS/GTC are capable of
providing comparable or better image resolution than acquired with STIS/HST and promise to
further populate the useful and insightful $M_{\rm bh}$-$\sigma_0$ diagram in
the future.  They of course additionally offer the ability to provide
two-dimensional velocity dispersion (and rotational) information and thereby
take us beyond the use of simple central velocity dispersion measurements 
and thereby better constrain the kinetic energy and mass of each galaxy or
bulge.

\section*{Acknowledgments}
%
%
I am happy to thank David Merritt and Stuart Wyithe for enjoyable
discussions pertaining to aspects of this work.
I acknowledges use of the HyperLeda database 
(http://leda.univ-lyon1.fr) 
and the NASA/IPAC Extragalactic Database (NED) which
is operated by the Jet Propulsion Laboratory, California Institute of
Technology, under contract with the National Aeronautics and Space
Administration.


\begin{table*}[ht]
\begin{center}
\caption{Fifty galaxies with direct SMBH mass measurements \label{Tab1}}
\begin{tabular}{llcclll}
\hline
 Galaxy &  Type  & core &  Dist.     & $\sigma_0$    &   $M_{\rm bh}$   &  Reference \\
        &        &      &  Mpc       & km s$^{-1}$ & $10^8 M_{\odot}$ &     \\
   1    &   2    &   3  &   4        &  5          &   6              &  7  \\
\hline
Circinus  &   Sb & ... &  2.8 [1]   &  75     &  $0.011^{+0.002}_{-0.002}$ & m-8 \\ 
Cygnus A  & E    & ... &  232 [2]   & 270     &  $25.0^{+7.0}_{-7.0}$      & g-9 \\
IC 2560   & SBb  & ... & 40.7 [2]   & 144 [6] & $^a0.044^{+0.044}_{-0.022}$ & m-10,11 \\
Milky Way & SBbc &  n  & 0.008 [3]  & 100     &  $0.037^{+0.002}_{-0.002}$ & p-12 \\
NGC 221   & S0   &  n  &  0.8        &  72     &  $0.025^{+0.005}_{-0.005}$ & s-13 \\
NGC 224   & Sb   &  n  &  0.8        & 170     &  $1.4^{+0.9}_{-0.3}$     & s-14,15 \\
NGC 821   & E    &  n  & 24.1        & 200     &  $0.85^{+0.35}_{-0.35}$  & s-16 \\
NGC 1023  & SB0  &  n  &  11.4       & 204     &  $0.44^{+0.05}_{-0.05}$  & s-17 \\
NGC 1300  & SBbc & ... & 20.7 [2]   & 229     &  $0.73^{+0.69}_{-0.35}$  & g-18 \\
NGC 1399  & E    &  y  & 20.0        & 329     &  $4.8^{+0.7}_{-0.7}$     & s-19,20 \\ 
NGC 2778  & SB0  &  n  & 22.9        & 162     &  $0.14^{+0.08}_{-0.09}$  & s-21 \\
NGC 2787  & SB0  & ... &   7.5       & 210     &  $0.41^{+0.04}_{-0.05}$  & g-22 \\
NGC 3031  & Sb   & ... &   3.9       & 162     &  $0.76^{+0.22}_{-0.11}$  & g-23 \\
NGC 3079  & SBcd & ... & 20.7 [2]   & 146     &  $^a0.024^{+0.024}_{-0.012}$ & m-24,25,26  \\
NGC 3115  & S0   &  n  &   9.7       & 252     &  $9.1^{+10.3}_{-2.8}$     & s-27 \\
NGC 3227  & SB   & ... & 20.3 [2]   & 133     &  $0.14^{+0.10}_{-0.06}$  & s-28 \\
NGC 3245  & S0   & ... &  20.9       & 210     &  $2.1^{+0.5}_{-0.5}$     & g-29 \\
NGC 3377  & E5   &  n  &  11.2       & 139     &  $0.8^{+0.05}_{-0.06}$   & s-21,30 \\ 
NGC 3379  & E    &  y  &  10.6       & 207     &  $1.4^{+2.7}_{-1.0}$     & s-31 \\
NGC 3384  & SB0  &  n  &  11.6       & 148     &  $0.16^{+0.01}_{-0.02}$  & s-21 \\
NGC 3608  & E2   &  y  &  22.9       & 192     &  $1.9^{+1.0}_{-0.6}$     & s-21 \\
NGC 3998  & S0   & ... &  14.1       & 305     &  $2.2^{+2.0}_{-1.7}$     & s-32 \\
NGC 4151  & SBab & ... &  20.0 [2]  & 156     &  $0.65^{+0.07}_{-0.07}$  & s-33 \\ 
NGC 4258  & SBbc & ... &   7.2 [4]  & 134     &  $0.39^{+0.01}_{-0.01}$  & m-34,4 \\
NGC 4261  & E2   &  y  &  31.6       & 309     &  $5.2^{+1.0}_{-1.1}$     & g-35 \\
NGC 4291  & E2   &  y  &  26.2       & 285     &  $3.1^{+0.8}_{-2.3}$     & s-21 \\
NGC 4342  & S0   & ... &  17.0 [5]   & 253     &  $3.3^{+1.9}_{-1.1}$     & s-36,37 \\
NGC 4374  & E    &  y  &  18.4       & 281     &  $4.64^{+3.46}_{-1.83}$  & g-38 \\
NGC 4459  &  S0  & ... &  16.1       & 178     &  $0.70^{+0.13}_{-0.13}$ & g-22 \\ 
NGC 4473  & E5   &  y  &  15.7       & 179     &  $1.1^{+0.4}_{-0.8}$     & s-21 \\
NGC 4486  & E0   &  y  &  16.1       & 332     &  $34^{+10}_{-10}$        & g-39 \\ 
NGC 4486a & E    & ... &  17.0 [5]   & 110 [32] & $0.13^{+0.08}_{-0.08}$  & s-40 \\
NGC 4564  & S0   &  n  &  15.0       & 157     &  $0.56^{+0.03}_{-0.08}$  & s-21 \\
NGC 4596  & SB0  & ... &  17.0 [5]   & 149     &  $0.79^{+0.38}_{-0.33}$  & g-22 \\ 
NGC 4649  & E1   &  y  &  16.8       & 335     &  $20^{+4}_{-6}$          & s-21 \\
NGC 4697  & E4   &  n  &  11.7       & 174     &  $1.7^{+0.2}_{-0.1}$     & s-21 \\ 
NGC 4945  & SBcd & ... &   3.8 [1]  & 100     &  $^a0.014^{0.014}_{-0.007}$ & m-41 \\
NGC 5077  & E    &  y  &  41.2 [2]  & 255     &  $7.4^{+4.7}_{-3.0}$     & g-42 \\
NGC 5128  & S0   &  n  &   3.8 [1]  & 120     &  $0.49^{+0.18}_{-0.11}$  & g-43,44 \\
NGC 5252  & S0   & ... & 103.5 [2]  & 190     &  $10.6^{+16.3}_{-5.0}$   & g-45 \\
NGC 5845  & E3   &  n  &  25.9       & 233     &  $2.4^{+0.4}_{-1.4}$     & s-21 \\
NGC 6251  & E    & ... & 104.6 [2]  & 311     &  $5.9^{+2.0}_{-2.0}$     & g-46 \\
NGC 7052  & E4   &  y  &  66.4 [2]  & 277     &  $3.7^{+2.6}_{-1.5}$     & g-47 \\
NGC 7582  & SBab & ... &  22.0 [2]  & 156     &  $0.55^{+0.26}_{-0.19}$  & g-48 \\
\hline
\end{tabular}
\end{center}
\end{table*}

\setcounter{table}{0}

\begin{table*}[ht]
\begin{center}
\caption{{\it cont.}}
\begin{tabular}{llcclll}
\hline
 Galaxy &  Type  & core &  Dist.     & $\sigma_0$    &   $M_{\rm bh}$   &  Reference \\
        &        &      &  Mpc       & km s$^{-1}$ & $10^8 M_{\odot}$ &     \\
   1    &   2    &   3  &   4        &  5          &   6              &  7  \\
\hline
\multicolumn{7}{c}{Preliminary SAURON/OASIS data} \\
NGC 2974  &  E   &  n  &  21.5       &  227    &  $1.7^{+0.3}_{-0.3}$     & s-7a \\
NGC 3414  &  S0  &  n  &  25.2       &  237    &  $2.5^{+0.4}_{-0.4}$     & s-7a \\
NGC 4552  &  S0  &  y  &  15.3       &  252    &  $4.8^{+0.8}_{-0.8}$     & s-7a \\ 
NGC 4621  &  E   &  n  &  18.3       &  225    &  $4.0^{+0.6}_{-0.6}$     & s-7a \\
NGC 5813  &  E   &  y  &  32.2       &  239    &  $7.0^{+1.1}_{-1.1}$     & s-7a \\
NGC 5846  &  E   & ... &  24.9       &  237    &  $11.0^{+2.0}_{-2.0}$    & s-7a \\
\hline
\end{tabular}
\end{center}
Unless otherwise specified, the distances have come from Tonry et al.\ (2001).
The distances from NED are 
the (Virgo + GA + Shapley)-corrected Hubble flow distances. 
The velocity dispersions have come from
HyperLeda\footnote{http://leda.univ-lyon1.fr} (Paturel et al.\ (2003) unless
otherwise noted. 
$M_{\rm bh}$ has been adjusted to the distance given in column~4. \\
$^a$ A factor of two uncertainty has been assigned to these SMBH masses. \\
References:
 1 = Karachentsev et al.\ (2007);
 2 = NED (http://nedwww.ipac.caltech.edu/);
 3 = Eisenhauer et al.\ (2003).
 4 = Herrnstein et al.\ (1999); 
 5 = Jerjen et al.\ (2004);
 6 = Cid Fernandes et al.\ (2004); 
 7 = Hu (2008); 
 7a = Preliminary values determined by Hu (2008) from Conf.\ Proc. figures of Cappellari et al.\ (2006, 2007); 
 8 = Greenhill et al.\ (2003a);  
 9 = Tadhunter et al.\ (2003);
10 = Ishihara et al.\ (2001);
11 = Nakai et al.\ (1998); 
12 = Ghez et al.\ (2005);
13 = Verolme et al.\ (2002);
14 = Bacon et al.\ (2001);
15 = Bender et al.\ (2005);
16 = Richstone et al.\ (2008);
17 = Bower et al.\ (2001);
18 = Atkinson et al.\ (2005);
19 = Houghton et al.\ (2006);
20 = Gebhardt et al.\ (2007); 
21 = Gebhardt et al.\ (2003);
22 = Sarzi et al.\ (2001);
23 = Devereux et al.\ (2003); 
24 = Trotter et al.\ (1998);
25 = Yamauchi et al.\ (2004); 
26 = Kondratko et al.\ (2005);
27 = Emsellem et al.\ (1999);
28 = Davies et al.\ (2006);          
29 = Barth et al.\ (2001); 
30 = Copin et al.\ (2004);
31 = Shapiro et al.\ (2006, stellar dynamical measurement); 
32 = De Francesco et al.\ (2006);
33 = Onken et al.\ (2007);           
34 = Miyoshi et al.\ (1995);
35 = Ferrarese et al.\ (1996);
36 = Cretton \& van den Bosch (1999);
37 = Valluri et al.\ (2004); 
38 = Maciejewski \& Binney (2001);
39 = Macchetto et al.\ (1997); 
40 = Nowak et al.\ (2007);
41 = Greenhill et al.\ (1997);
42 = De Francesco et al.\ (2008); 
43 = Marconi et al.\ (2006);
44 = Neumayer et al.\ (2007); 
45 = Capetti et al.\ (2005);
46 = Ferrarese \& Ford (1999);
47 = van der Marel \& van den Bosch (1998);
48 = Wold et al.\ (2006). 

\end{table*}

\clearpage

\begin{table*}[ht]
\begin{center}
\caption{Additional Galaxies \label{Tab2}}
\begin{tabular}{llcclll}
\hline
 Galaxy & Type  &  core & Dist. & $\sigma_0$    & $M_{\rm bh}$        &   Reference \& comment \\
        &       &       &  Mpc  & km s$^{-1}$ & $10^8 M_{\odot}$ &      \\
   1    &   2   &   3   &  4    &      5      &      6              &    7 \\
\hline 
\multicolumn{7}{c}{Twenty six galaxies with somewhat uncertain $M_{\rm bh}$ values} \\
Abell 1836 &  BCG & ... &  157      & ... &  $48^{+8}_{-7}$             & g-4, no refereed publication \\
A2052/UGC 9799 & BCG & y? & 155  & 234 &  $< 73$                  & g-4, no refereed publication \\
A3565/IC 4296 & BCG & ... & 40.7  & 336 & $13^{+3}_{-4}$           & g-4, no refereed publication \\
ESO 269-G012 & S0 & ... &  59.6  & ... &  $0.01 - 0.1$               & m-5, Maser, modelling uncertain \\
IC 1459    &  E3  &  y  &  29.2 [1] & 306 &  $3-36$                     & g,s-6, gas/stellar dynamics differ \\
NGC 1068   &  Sb  & ... &  15.2     & 151 &  $0.084^{+0.003}_{-0.003}$  & m-7, Maser, modelling uncertain \\
NGC 1386   &  SB0 & ... &  16.5 [1] & 166 & $^a0.012^{+0.012}_{-0.006}$ & m-8, Maser, modelling uncertain \\
NGC 2639   & SBa  & ... &  49.6     & 198 &  $0.16 (r/0.1 {\rm pc})^2$  & m-9, Maser, modelling uncertain \\
NGC 2748   & Sbc  & ... &  25.1     &  92 &  $0.48^{+0.38}_{-0.39}$     & g-10 \& 11, Dust an issue \\ 
NGC 2960   &  Sa  & ... &  72.8     & ... &  $0.12^{+0.03}_{-0.03}$     & m-12, Maser, modelling uncertain \\ 
NGC 3393   & SBab & ... &  55.2     & 197 &  $0.31^{+0.02}_{-0.02}$     & m-13, Maser, modelling uncertain \\
NGC 4041   &  Sbc & ... &  23.3     &  95 [14] & $<0.24$                & m-14, disk dynamically decoupled(?) \\
NGC 4303   & SBbc & ... &  16.1 [2] & 109 &  $0.006 - 0.160$            & g-15, poorly known disk inclination \\
NGC 4350   &  S0  & ... &  17.0 [3] & 181 &  1.5-9.8                    & g,s-16, high $M_{\rm bh}/M_{\rm bulge}$ \\
NGC 4435   & SB0  & ... &  14.0     & 157 &  $<0.075$                   & g-17, possibly no black hole \\ 
NGC 4486B  &  cE  &  n  &  17.0 [3] & 169 &  $6^{+3}_{-2}$              & s-18, $M_{\rm bh}/M_{\rm bulge} = 0.09$ \\
NGC 4594   &  Sa  & ... &   9.8 [1] & 240 &  $1.7 - 17$                 & s-19, no 3-integral model \\
NGC 4742   & E4   &  n  &  15.5 [1] & 109 &  $0.14^{+0.04}_{-0.05}$     & s-20, no refereed publication \\
NGC 5055   &  Sbc & ... &   8.7     & 101 &  $8.5^{+1.9}_{-1.9}$        & g-21, possibly no black hole \\
NGC 5495   &  Sb  & ... & 103       & ... &  $^a0.12^{+0.12}_{-0.06}$   & m-13, Maser, modelling uncertain \\
NGC 5793   &  Sb  & ... &  53.3     & ... &  $^a\sim0.1^{+0.1}_{-0.1}$  & m-22, Maser, modelling uncertain \\
NGC 6926   & SBbc & ... &  84.0     & ... &  $0.01 - 0.1$               & m-5, Maser, modelling uncertain \\
NGC 7332   &  S0  &  n  &  23.0 [1] & 135 &  $0.13^{+0.06}_{-0.05}$     & s-23, no refereed publication \\
NGC 7457   & S0   &  n  &  13.2 [1] &  69 &  $0.035^{+0.011}_{-0.014}$  & s-24, AGN/NC distinction blurred \\
NGC 7469   & SBa  & ... &  67.0     & 153 [25] &  $<0.5$             & g-26, possibly no black hole \\
UGC 3789   &  Sab & ... &  48.4     & ... &  $^a0.09^{+0.09}_{-0.04}$   & m-27, Maser, modelling uncertain \\
\multicolumn{7}{c}{Eight Intermediate Mass Black Hole candidates}  \\
G1         &   GC & ... &  0.8  [1a] & 25       &  $1.8^{+0.5}_{-0.5} \times 10^{-4}$  & s-34, but see s-35 \\
M15        &   GC & ... &  0.01 [28] & 14    &  $1.7^{+2.7}_{-1.7} \times 10^{-5}$  & s-36, consistent with no IMBH (s-37) \\
M33        & Scd  &  n  &  0.8  [29] & 24    &  $<3 \times 10^{-5}$                 & s-38 \& 39, consistent with no IMBH \\
MGG-11     & Irr  &  n  &  3.6  [30] & 11.4 [30] &  $1.0^{+4.0}_{-0.8}\times10^3$       & x-40 \\
NGC 205    &   E5 &  n  &  0.82 [31] & 23    &  $<2.2 \times 10^{-4}$               & s-41, consistent with no IMBH \\   
NGC 4395   & Sm   &  n  &  4.3  [32] & 20-35 &  $^b10^{-4} - 10^{-3}$               & 42, consistent with no IMBH \\
$\omega$ Cen & GC & ... & 0.0048 [33] & 20-23 & $4.0^{+0.75}_{-1.0} \times 10^{-4}$ & s-43, alternatives not ruled out \\
Pox 52     & dE   &  n  & 98.8          & 36 & $^{a,b}3.2^{+1.0}_{-1.0}\times 10^{-3}$ & 44,45 indirect estimates \\
\hline 
\end{tabular}
\end{center}
Unless otherwise specified, the distances have come from NED, and are 
the (Virgo + GA + Shapley)-corrected Hubble flow distances.  
The velocity dispersions have come from
HyperLeda\footnote{http://leda.univ-lyon1.fr} (Paturel et al.\ 2003) unless
noted otherwise. 
$M_{\rm bh}$ has been adjusted to the distance given in column~4. \\
$^a$ A factor of two uncertainty has been assigned to these BH masses. \\
$^b$ BH mass obtained from the line width-luminosity-mass relation
rather directly probing resolved kinematics about the BH.\\
References:
1 = Tonry et al.\ (2001); 
1a = the Tonry et al.\ (2001) distance to NGC~224 (M31) is used;
2 = Ferrarese et al.\ (1996);
3 = Jerjen et al.\ (2004);
4 = Dalla Bont\`a et al.\ (2006); 
5 = Greenhill et al.\ (2003b);  
6 = Cappellari et al.\ (2002);
7 = Lodato \& Bertin (2003);
8 = Braatz et al.\ (1997); 
9 = Wilson et al.\ (1995); 
10 = Atkinson et al.\ (2005);
11 = Hu (2008); 
12 = Henkel et al.\ (2002); 
13 = Kondratko et al.\ (2006):
14 = Marconi et al.\ (2003); 
15 = Pastorini et al.\ (2007);
16 = Pignatelli et al.\ (2001);
17 = Coccato et al.\ (2006); 
18 = Kormendy et al.\ (1997); 
19 = Kormendy (1988); 
20 = Tremaine et al.\ (2002);
21 = Blais-Ouellette et al.\ (2004);
22 = Hagiwara et al.\ (2001); 
23 = H\"aring \& Rix (2004); 
24 = Gebhardt et al.\ (2003);
25 = Peterson et al.\ (2004);
26 = Hicks \& Malkan (2008); 
27 = Braatz et al.\ (2008); 
28 = Harris (1996);
29 = Argon et al.\ (2004); 
30 = McCrady et al.\ (2003); 
31 = McConnachie et al.\ (2005); 
32 = Thim et al.\ (2004);  
33 = van de Ven et al.\ (2006); 
34 = Gebhardt et al.\ (2005); 
35 = Baumgardt et al.\ (2003b); 
36 = Gerssen et al.\ (2003); 
37 = Baumgardt et al.\ (2003a); 
38 = Gebhardt et al.\ (2001);
39 = Merritt et al.\ (2001); 
40 = Patruno et al.\ (2006); 
41 = Valluri et al.\ (2005);  
42 = Filippenko \& Ho (2003);  
43 = Noyola et al.\ (2008); 
44 = Barth et al.\ (2004); 
45 = Thornton et al.\ (2008).
\end{table*}

\end{document}